\documentclass[reprint,amsmath,amssymb,aip,jap]{revtex4-1}
\usepackage[utf8]{inputenc}
\usepackage{mathrsfs}
\usepackage{amsmath}
\usepackage{bm}
\usepackage{amsfonts}
\usepackage{amssymb}
\usepackage{amsthm}
\usepackage{color}
\usepackage{graphicx}
\graphicspath{{./images/}}
\usepackage{geometry}
\usepackage{hyperref}
\usepackage{ulem}
\usepackage{epstopdf}
\usepackage{graphicx}
\hypersetup{
     unicode=false,
     pdftoolbar=true,
     pdfmenubar=true,
     pdffitwindow=false,     
     pdfstartview={FitH},    
     pdftitle={My title},    
     pdfauthor={Author},     
     pdfsubject={Subject},   
     pdfcreator={Creator},   
     pdfproducer={Producer}, 
     pdfkeywords={keyword1} {key2} {key3}, 
     pdfnewwindow=true,      
     colorlinks=false,       
     linkcolor=red,          
     citecolor=green,        
     filecolor=magenta,      
     urlcolor=cyan           
}
\begin{document}
\title{Current-induced skyrmion motion on magnetic nanotubes}
\author{Xiaofan Wang}
\author{X. S. Wang}
\email[Corresponding author: ]{justicewxs@connect.ust.hk}
\author{C. Wang}
\author{Huanhuan Yang}
\author{Yunshan Cao}
\author{Peng Yan}
\email[Corresponding author: ]{yan@uestc.edu.cn}
\affiliation{School of Electronic Science and Engineering and State Key Laboratory of Electronic Thin Films and Integrated Devices, University of Electronic Science and Technology of China, Chengdu 610054, China}

\begin{abstract}
Magnetic skyrmions are believed to be the promising candidate of information carriers in spintronics.
However, the skyrmion Hall effect due to the nontrivial topology of skyrmions can induce a skyrmion accumulation or even annihilation at the edge of the devices, which hinders the real-world applications of skyrmions. In this work, we theoretically investigate the current-driven skyrmion motion on magnetic nanotubes which can be regarded as ``edgeless" in the tangential direction. By performing micromagnetic simulations, we find that the skyrmion motion exhibits a helical trajectory on the nanotube, with its axial propagation velocity proportional to the current density. Interestingly, the skyrmion's annular speed increases with the increase of the thickness of the nanotube. A simple explanation is presented. Since the tube is edgeless for the tangential skyrmion motion, a stable skyrmion propagation can survive in the presence of a very large current density without any annihilation or accumulation. Our results provide a new route to overcome the edge effect in planar geometries.
\end{abstract}

\maketitle

Ever since its experimental discovery,\cite{M2009Skyrmion} the magnetic skyrmion, a chiral
quasiparticle,\cite{Nagaosa2013Topological,Sampaio2013Nucleation} has been an active research
area in condensed matter physics because of not only the potential for future spintronic
applications such as skyrmion racetrack memories \cite{Fert2013,Kang2016,Jmuller2017} and
logic devices,\cite{skyrDevice1,skyrDevice2} but also the fundamental interests.\cite{gyration1,gyration2,YangOE2018,YangPRL2018,LiPRB2018}
In chiral magnets, skyrmions can be stabilized by the Dzyaloshinskii-Moriya interaction (DMI) of two types:\cite{Rossler2006,M2009Skyrmion,Yu2010,Onose2012,Park2014,Du2015,Heinze2011,Romming2013,Jiang2015,Krause2016}
the bulk DMI and the interfacial one. The bulk DMI typically exists in noncentrosymmetric magnets, and can support the formation of Bloch-type (vortex-like) skyrmions,\cite{M2009Skyrmion,Yu2010,Onose2012,Park2014,Du2015} while the latter one usually exists in inversion-symmetry breaking thin films, and can give rise to N\'{e}el-type (hedgehog-like) skyrmions.\cite{Heinze2011,Romming2013,Jiang2015,Krause2016}

\begin{figure}[htbp!]
  \centering
  \includegraphics[width=8.5cm]{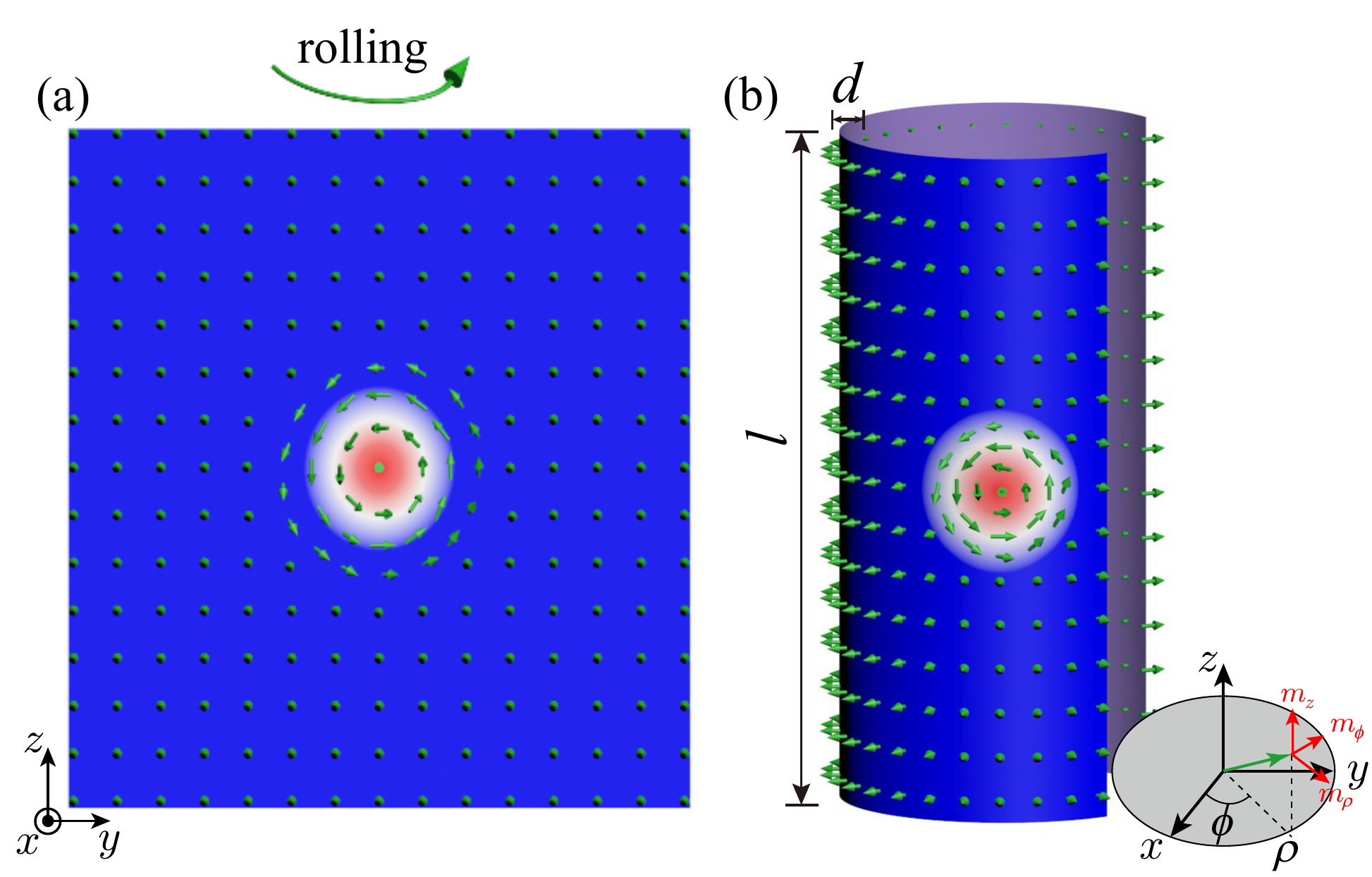}\\
  \caption{(a) Schematic illustration of a Bloch-type skyrmion in a planar film. Green arrows refer to the local spin directions. (b) Skyrmion on a nanotube by rolling up (a). Colors refer to the $\rho$-component of the magnetization. The coordinate system is defined
  in the inset.}\label{fig1}
\end{figure}

Several methods have been proposed to drive the skyrmion motion, such as spin-polarized currents,\cite{Tomasello2014A} microwaves,\cite{microwave} and thermal gradients,\cite{Kong2013Dynamics} to name a few. However, when the skyrmion is driven by an
in-plane current via the spin transfer torque, the trajectory of its motion deviates from the current direction
due to the intrinsic skyrmion Hall effect.\cite{Nagaosa2013Topological,Sampaio2013Nucleation,Michael1996,Iwasaki2013,Iwasaki2013U,Wanjun2017}
Furthermore, there exists a threshold current density above which skyrmions can annihilate at the film edge.\cite{Yoo2017}
This edge effect strongly limits the speed of skyrmion propagation which is of vital importance for real applications.
Several solutions have been proposed to overcome this problem. Zhang \textit{et al.} proposed an antiferromagnetically
exchange-coupled bilayer system, where the skyrmions move straightly along the current direction.\cite{Zhang2016}
Upadhyaya \textit{et al.} showed that the skyrmion can be guided in a desired trajectory by applying electric fields in a certain pattern.\cite{Upadhyaya2015} More recently, Yang \textit{et al.} discovered a novel twisted skyrmion state at the boundary of two antiparallel magnetic domains coupled antiferromagnetically, through which skyrmions with opposite polarities can transform mutually.\cite{Yang2018} Under proper conditions, the domain boundary can also act as a reconfigurable channel for skyrmion propagations.\cite{Yang2018} All these proposals were aiming to eliminate the skyrmion Hall effect in a planar geometry. Different from a planar strip with two edges, a closed curved geometry, e.g., magnetic spheres and/or cylinders, can be edgeless. In such geometries, the skyrmion cannot vanish at edges any more, even in the presence of the skyrmion Hall effect.
This fact motivates us to consider the skyrmion motion on a nanotube that a planar strip is rolled up, as shown in Fig. \ref{fig1}.

In this work, we show, via micromagnetic simulations, that the skyrmion can be created on magnetic nanotubes and the skyrmion motion exhibits a helical trajectory when it is driven by an electric current along the tube. Further, we demontrate that the skyrmion can travel over arbitrarily long distances in the presence of a very large current density since the nanotube geometry are edgeless. The skyrmion's annular speed increases with the increase of the thickness of the nanotube, which is different from the case in planar geometry.

We consider the magnetic energy density in a nanotube,
\begin{equation}
\mathcal{E}=A_\text{ex}\left|\nabla\mathbf{m}\right|^2+D\mathbf{m}\cdot(\nabla\times\mathbf{m})-K(\mathbf{m}\cdot\hat{\rho})^2+\mathcal{E}_{\text{DDI}},
\label{energy}
\end{equation}
where $\mathbf{m}$ is the unit magnetization vector with a saturation magnetization $M_s$,
$A_\text{ex}$ is the ferromagnetic exchange constant, $D$ is the bulk Dzyaloshinskii-Moriya interaction (DMI) strength, $K>0$ is the easy-normal anisotropy constant along
$\hat{\rho}$ direction, and $\mathcal{E}_{\text{DDI}}$ is the energy density of dipole-dipole interaction.
$\left|\nabla\mathbf{m}\right|^2$ is short for $\left|\nabla m_x\right|^2+\left|\nabla m_y\right|^2
+\left|\nabla m_z\right|^2$.

To study the current-driven magnetization dynamics, we solve the Landau-Lifshitz-Gilbert equation with the spin transfer torque $\bm{\tau}_{\rm stt}$ associated with the electric current flowing along the tube,\cite{ZhangLi2004,Vansteenkiste2014}
\begin{equation}
\partial_t \mathbf{m} = -\gamma\mathbf{m} \times \mathbf{H}_{\rm{eff}} + \alpha \mathbf{m} \times \partial_t \bf{m}+\bm{\tau}_{\rm stt}.
\label{LLG}
\end{equation}
Here $\gamma$ is the gyromagnetic ratio, $\alpha$ is the Gilbert damping constant, and $\mathbf{H}_{\rm{eff}}=-\frac{\delta \mathcal{E}}{M_s\delta\mathbf{m}}$ is the effective field. The spin transfer torque $\bm{\tau}_{\rm stt}$ can be written as
\begin{equation}
\bm{\tau}_{\rm stt}=-(\bf{v}_s \cdot \nabla)\bf{m}+\beta \bf{m} \times (\bf{v}_s \cdot \nabla)\bf{m},
\end{equation}
where $\mathbf{v}_s=-\mu_{B}p\mathbf{j}_e/[eM_s(1+\beta^2)]$ is a vector with dimension of velocity and parallel
to the spin-polarized current density $\mathbf{j}_e$, $e$ is the electronic charge, $p$ is the polarization rate of the current, $\mu_{B}$ is the Bohr magneton,
and $\beta$ is the dimensionless parameter describing the degree of non-adiabaticity.\cite{ZhangLi2004}

To visualize the skyrmion motion on magnetic nanotubes, we performed micromagnetic simulations by employing the MuMax3 package.\cite{Vansteenkiste2014} The nanotube for numerical study is defined by fixed outer radius $R=50\rm~nm$,
various thickness $d = 10-40\rm~nm$, and length $l=600~\rm nm$.
The mesh size of $\rm 2 \times2 \times 2~nm^{3}$ is used in our simulations.
The magnetic nanotubes are assumed to be made of FeGe and the following material parameters are used:\cite{parameter} exchange stiffness $ A\rm_{ex}=8.78 ~pJ/m$, saturation magnetization $M_s =\rm~1.1\times 10^5 \rm~A/m$, bulk DMI parameter $D$ varying from $\rm0.8~mJ/m^2$ to $\rm1.5~mJ/m^2$, easy-normal anisotropy parameter $K=2\times 10^5  \rm~J/m^3$, and Gilbert damping constant $\alpha=0.1$. For the spin transfer torque, we assume $p=0.5$ and $\beta=0.5$.
Figure \ref{fig1} schematically shows a Bloch-type skyrmion (a) in a planar film and (b) in a nanotube. The coordinate system is
shown at the lower right corner of Fig. \ref{fig1}(b). $\rho$, $\phi$ and $z$ represent the radial, tangential, and axial coordinates, respectively.
The origin is set to be the center of the tube.

\begin{figure}
  \centering
  \includegraphics[width=8.5cm]{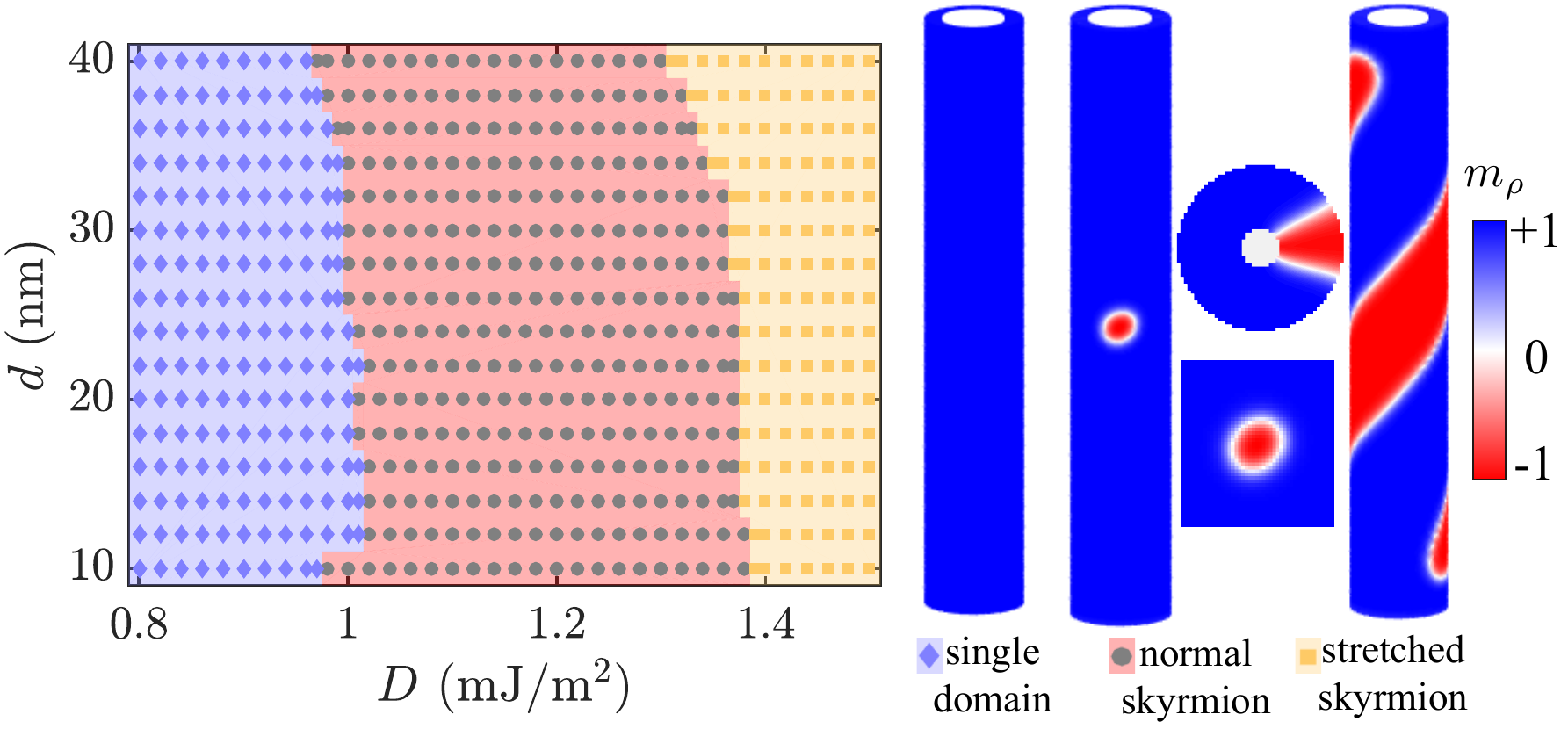}\\
  \caption{The magnetic phase diagram as a function of $D$ and $d$. Three phases are identified: single domain (rhombuses), ordinary isolated skyrmion (circles), and stretched skyrmion (squares). Representative magnetization profiles are shown at the right panel. From left to right: single domain, ordinary isolated skyrmion with both the sectional view (upper) and the side view (lower), and stretched skyrmion.}\label{PD}
\end{figure}

Firstly, in order to see how a skyrmion can exist on magnetic nanotubes, we numerically calculate the phase
diagram by tuning parameters $D$ and $d$.
We initially set $m_\rho=-1$ on the intersection of a 20-nm-diameter cylinder along the half-line $\phi=0$, $z=0$ and the nanotube,
 and set $m_\rho=1$ in the rest part, and then relaxed the
system from the initial state by minimizing the total energy.
Numerical results are shown in Fig. \ref{PD}, in which three phases are identified: single domain of $m_\rho=1$ (rhombuses),
ordinary isolated skyrmion (circles), and stretched skyrmion (squares) where the skyrmion is elongated like a spiral reaching the ends of nanotube. The typical magnetization profiles are shown next to the phase diagram.
When $D$ is small, the stable state is a single domain. The phase boundary
between the single-domain phase and the ordinary skyrmion phase is mesh-size-dependent. That is because when $D$ is small, the skyrmion size is also small \cite{Beach2018,wxs2018} so that the 2-nm mesh is not small enough to mimic the continuous model. To justify this, we have tested that
when the mesh size is $1$ nm, the stable state becomes an ordinary isolated skyrmion for $D=\rm1~mJ/m^2$.
For an intermediate $D$, an isolated skyrmion can exist. The sectional view cut in $xy$ plane
(upper) and the front view after expanding the tube into a plane (lower) are shown
for $D=1.2$ mJ/m$^2$ and $d=40$ nm.
From the inner surface to the outer surface, the skyrmion size is getting larger. The skyrmion size (or radius) is almost linearly dependent on $\rho$.
It can also be observed that the skyrmion is tilted to the right [see also Fig. \ref{fig4}(a) below]. This is because of the effective DMI induced by the
curvature of the tube, which will be explained later. When $D$ is large enough, the stable state is a stretched skyrmion.
The stretched skyrmion forms a right-handed spiral on the tube when
$D>0$ as shown in the figure, or a left-handed spiral when $D<0$ (not shown). The phase boundary between the ordinary skyrmion phase and the stretched skyrmion phase depends on
the diameter of the tube, similar to the well known fact that the confinement effect of the sample boundary is important
in stabilization of the skyrmion in planar films.\cite{Thiaville2013} In planar films,
the upper limit of $D$ for the existence of an isolated skyrmion is larger in a small sample than that
in an infinite film.\cite{Thiaville2013,wxs2018} However, the upper limit of $D$ here in the nanotube is smaller than that
in an infinite film, probably because the skyrmion size is larger than the diameter of the nanotube.
The skyrmion size increases with $d$ due to the demagnetizing field, similar to the case in the planar geometry.\cite{Beach2018,wxs2018}

\begin{figure}
  \centering
  \includegraphics[width=8.5cm]{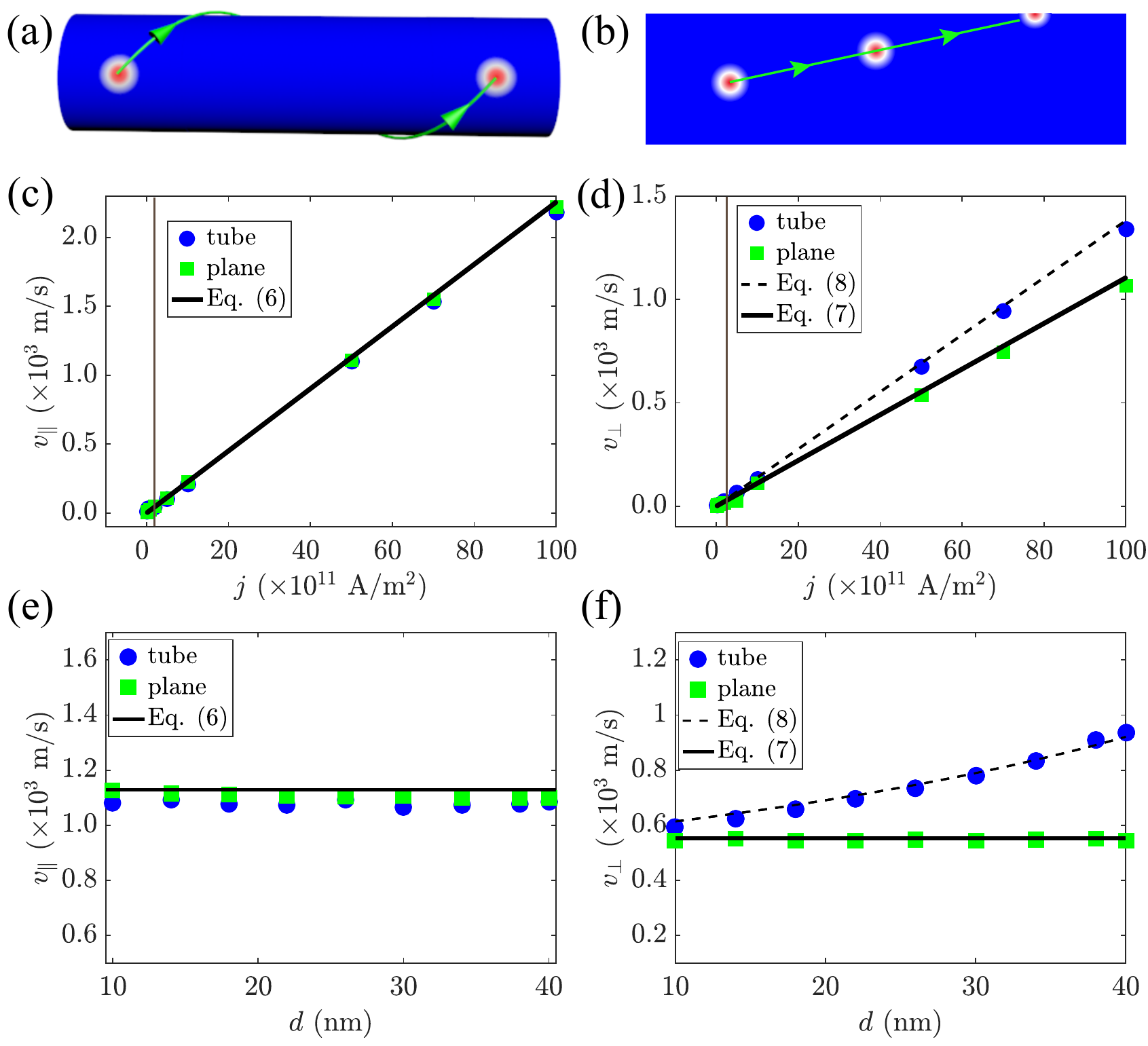}\\
  \caption{The trajectory of current-driven skyrmion motion in the nanotube (a) and the planar film (b). The current-dependence of skyrmion velocity, $v_\parallel$(c) and $v_\perp$(d), in which $d$ is fixed to $20~\rm nm$ and the vertical grey line denotes the annihilation of skyrmion at edges for the case of planar geometry, where the velocity is calculated before the annihilation of skyrmions. (e)-(f) The $d$-dependence of $v_\parallel$ and $v_\perp$ in both planar and tube geometries, in which $j$ is fixed to $5\times10^{12}\rm~A/m^2$. Symbols are numerical results. Solid lines are analytical results obtained from Eqs. (\ref{thiele1}) and (\ref{thiele2}) for planar films and dashed curves are from Eq. \eqref{tubehalf}.}\label{move}
\end{figure}

\begin{figure}
  \centering
  \includegraphics[width=8.5cm]{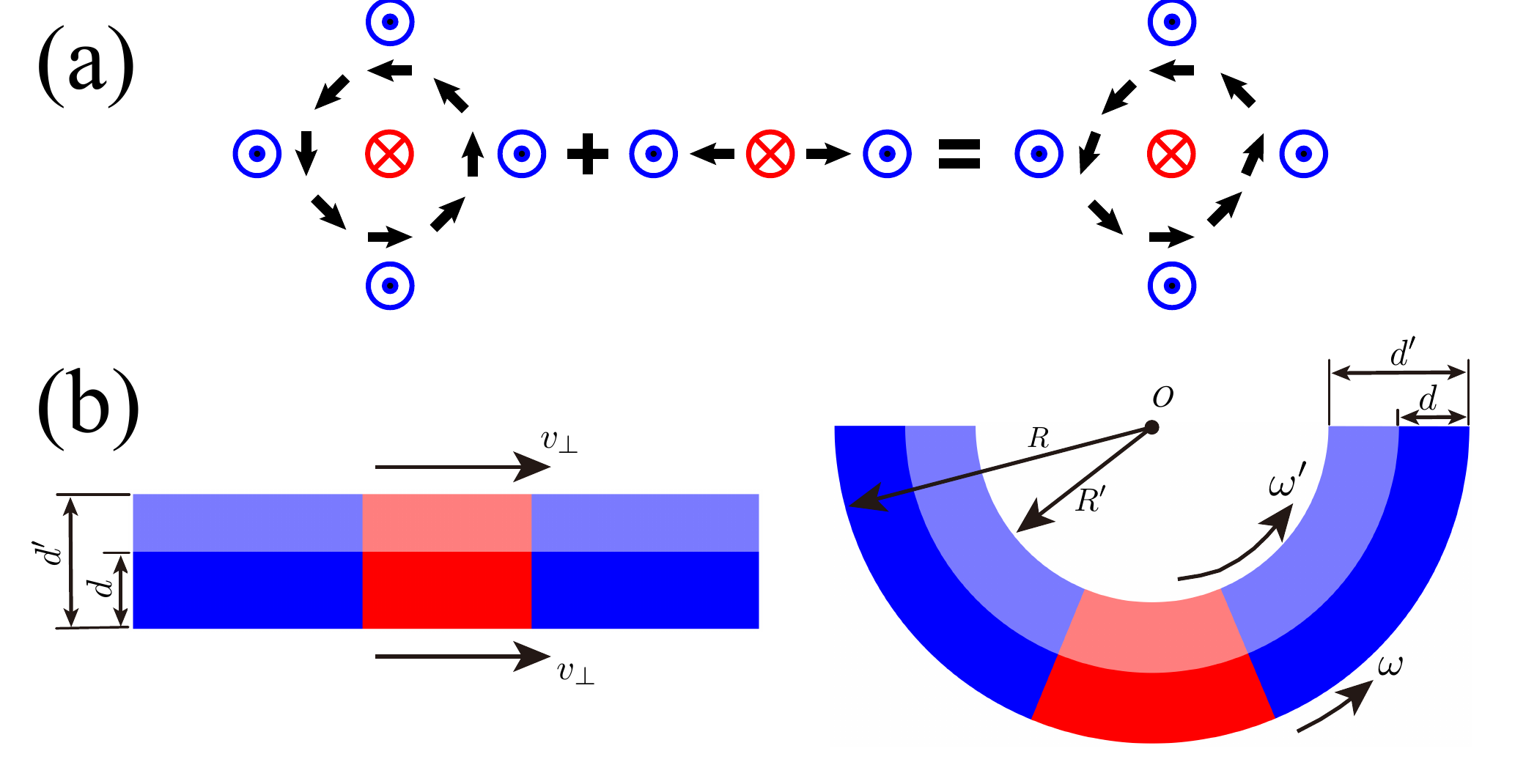}\\
  \caption{(a) Schematic diagram to explain the tilting of the skyrmion as well as the rotation sense of the elongated
  skyrmion. (b) Schematic picture to interpret why the linear velocities are different in the planar and nanotube geometries. The left panel shows the sectional view of the skyrmion motion in the planar geometry, in which $v_\perp$ stays unchanged when the thickness of planar geometry varies from $d$ to $d'$. The right panel shows the sectional view of the skyrmion motion in the nanotube geometry, in which the skyrmion's angular velocity inside the nanotube is faster than the one outside the nanotube, if layers are assumed to be decoupled.}\label{fig4}
\end{figure}

To investigate the current-driven skyrmion dynamics in infinite long nanotubes, we employ the periodic boundary condition in $z$-direction.
The DMI is fixed to be $D=1.2$ mJ/m$^2$. A typical current-driven skyrmion motion in the nanotube is plotted in Fig. \ref{move}(a) for $d=20~\rm nm$.
A skyrmion is initially created at one end of the nanotube. Then we inject an electric current along the axis of the nanotube. The current exerts spin transfer torques on the magnetization texture. Similar to the skyrmion motion in
planar film,\cite{Nagaosa2013Topological,Sampaio2013Nucleation,Michael1996} the skyrmion moves not only along the current direction, but also in the tangential direction at the same time because of the skyrmion Hall effect.
As a result, the skyrmion trajectory follows a helical curve, as shown in the green line of Fig. \ref{move}(a).
As a comparison, the trajectory of skyrmion motion in a planar film of the same thickness is shown in Fig. \ref{move}(b).
Due to the skyrmion Hall effect, the skyrmion will annihilate at the edge when the current density is large enough.
To see more details, we calculate the instantaneous velocity of the skyrmions from
the simulation results.
Considering the outer surface only, the skyrmion velocity $\mathbf{v}$ has two components:
the component parallel to the direction of the electric current, $v_\parallel = \mathbf{v} \cdot \hat{z}$,
and a perpendicular one, $v_\perp = \mathbf{v} \cdot \hat{\mathrm{\phi}}=R\omega$, where $\omega$ is the
skyrmion's ``annular speed".
Figures \ref{move}(c) and (d) show the current dependence of $v_\parallel$ and $v_\perp$, respectively. The two velocity components of the
skyrmion in planar film are also plotted as comparisons (It is noted that in planar film the velocity is measured before the skyrmion annihilation at the edge).
As the injected current density $j$ varies from $1\times 10^{10}\rm~A/m^2$ to $1\times 10^{13}\rm~A/m^2$, both $v_\parallel$ and $v_\perp$ are
proportional to $j$. In planar film, the skyrmion annihilates when $j$ is larger than $2\times 10^{11}$ A/m$^2$, as shown by the vertical grey lines in
Figs. \ref{move}(c) and (d). The corresponding longitudinal speed is no more than $100$ m/s.
However, because the tube is closed in the tangential direction, the skyrmion does not vanish even at a very large
current density.
The $v_\parallel$ can reach $\sim 2000~\rm m/s$ when we inject a very large $j=1\times 10^{13}\rm~A/m^2$.
So the skyrmion motion in nanotube geometries possesses the advantages that a stable skyrmion propagation can survive in the presence of a very large current density and the propagation speed can be very fast.

We then fix the current density $j=5\times 10^{12}\rm~A/m^2$, and investigate the $d$-dependence
of skyrmion velocity at the outer surface of the nanotube.
The outer radius of the nanotube is still $R=50$ nm as above. Figures \ref{move}(e) and (f) show the $d$-dependence of
$v_\parallel$ and $v_\perp$, respectively. For different thickness $d$, $v_\parallel$ is almost a constant [blue circles in Fig. \ref{move}(e)],
and shows no apparent difference in comparison with that in planar films [green squares in Fig. \ref{move}(e)].
However, $v_\perp$ increases with $d$ [blue circles in Fig. \ref{move}(f)], which is different from that in planar films,
where $v_\perp$ stays unchanged when $d$ increases [green squares in Fig. \ref{move}(f)]. See supplementary material MOVIE 1 for
skyrmion motions in different thicknesses.

To better understand the numerical findings, we first consider the effect of the curvature of the nanotube.
Suppose the nanotube is constructed by many coaxial thin layers of tubes, with
their thicknesses much smaller than their radii $\rho$.
For each layer, we can express the energy density in local coordinates on the
outer surface of the nanotube constructed
by basis vectors $(\hat{\rho},\hat{\phi},\hat{z})$.\cite{Gaididei2014,Kravchuk2016} Intuitively,
this means to expand the tube into a planar film. To be more clear in comparison with
a planar film, we rename the local basis vectors as $\hat{\rho}\rightarrow \hat{x^\prime}$,
$\hat{\phi}\rightarrow \hat{y^\prime}$, and $\hat{z}\rightarrow \hat{z^\prime}$.
In the local coordinates, the energy density is
\begin{multline}\label{cur}
  \mathcal{E}=A_\text{ex}\left|\nabla^\prime \mathbf{m}\right|^2+D\mathbf{m}\cdot(\nabla^\prime\times\mathbf{m})
  -K m_{x^\prime}^2+\mathcal{E}_\text{DDI}\\
  +A_\text{ex}\left[\frac{1}{\rho^2}+\frac{2}{\rho}\left(m_{x^\prime}
  \frac{\partial m_{y^\prime}}{\partial y^\prime}-m_{y^\prime}
  \frac{\partial m_{x^\prime}}{\partial y^\prime}\right)\right]\\
  -\frac{A_\text{ex}}{\rho^2} m_{z^\prime}^2+\frac{D}{\rho}m_{y^\prime}m_{z^\prime},
\end{multline}
where $\nabla^\prime$ denotes the derivatives in primed coordinates.
Compared to a planar film, the curvature induces three extra terms.
The first term $2A_\text{ex}\left(m_{x^\prime}
  \partial_{y^\prime} m_{y^\prime}-m_{y^\prime}
  \partial_{y^\prime} m_{x^\prime}\right)/\rho$
  comes from the exchange interaction. For a fixed $\rho$, this term has the
same mathematical form as an interfacial DMI along $\hat{z}$ direction
\cite{Wu2017} (the constant term $A_\text{ex}/\rho^2$ does not affect the magnetization profile). A left-handed N\'{e}el wall \cite{Heide2008,Beach2013} is thus preferred along
$\hat{z}$ direction. However, it is known that the bulk DMI prefers a Bloch skyrmion.
The superposition of this two effects therefore gives a skyrmion tilting to the right,
as schematically plotted in the Fig. \ref{fig4}(a), which is consistent with the numerical
observation shown in Fig. \ref{PD}. When the sign of $D$ is reversed,
the rotation direction of the Bloch skyrmion is also switched, so the orientation
of the tilting is flipped consequently. The stretched skyrmion also
grows with a certain tilting direction for the same reason. The second term is an effective easy-axis anisotropy along
$z^{(\prime)}$ direction, due to which the skyrmion is stretched
along the axial direction of the tube. Note that $A_\text{ex}/2\rho^2$
is smaller than $K$ for the parameters we used so that
the easy-normal anisotropy $K$ still dominates. The third term $Dm_{y^\prime}m_{z^\prime}/\rho$ comes from the DMI, and it prefers that
$m_{z^\prime}$ and $m_{y^\prime}$ have opposite signs, which competes with the first term.
We note that the first $A_\text{ex}$-term is approximately
$2A_\text{ex}m_{y^\prime}m_{z^\prime}/(\rho w)$ where $w\approx \pi D/(4K)$ is
the skyrmion wall width.\cite{Beach2018,wxs2018} For parameters used in
the simulations, $2A_\text{ex}/w$ is much larger than $D$. Thus, the first $A_\text{ex}$-term
dominates over the $D$-term, and the contribution from the latter term can be safely ignored.

Below, we analytically understand the motion of the skyrmion on the nanotube. Let's first consider a planar film, in which the skyrmion follows
the Thiele's equation,\cite{Thiele1973,Zhang2016,Tomasello2014A,Iwasaki2013U,Yoo2017,Sampaio2013Nucleation}
\begin{equation}
\mathbf{G} \times (\mathbf{v}-\mathbf{v_s})+\mathcal{D}(\alpha \mathbf{v} -\beta \mathbf{v}_s)=0 \label{thiele},
\end{equation}
where $\mathbf{G}=4\pi Q\hat{x}$ is the gyrovector where $Q=\pm1$ is the skyrmion number,
$\mathcal{D}_{ij}=\int \partial_i\mathbf{m}\cdot\partial_j
{\mathbf{m}}dydz$ is the dissipation tensor, and $\mathbf{v}$ is the skyrmion velocity.
For a rotationally symmetric skyrmion, $\mathcal{D}$ degenerates to a scalar that can be
calculated from the numerical results.
In our simulations, the electric current is applied only along $\hat{z}$ direction, and the Thiele's equation can be
easily solved. The parallel component ($z$-component) and the perpendicular component ($y$-component) are
\begin{gather}
v_\parallel=\left[\frac{\beta}{\alpha}+\frac{G^2}{\alpha}\frac{\alpha-\beta}{G^2+(\alpha\mathcal{D})^2}\right]v_s \label{thiele1},\\
v_\perp=\frac{(\alpha-\beta)G\mathcal{D}}{G^2+(\alpha\mathcal{D})^2}v_s\label{thiele2}.
\end{gather}
As shown by solid lines in Figs. \ref{move}(c)-(f), our analytical results obtained from Eqs. (\ref{thiele1})
and (\ref{thiele2}) are consistent with the micromagnetic simulations.

In the nanotube, for each layer of radius $\rho$ expanded to a planar film, the skyrmion still
follows the Thiele's equation. The gyrovector $\mathbf{G}$ does not depend on $\rho$.
Strictly speaking, the dissipation tensor $\mathcal{D}$ is no longer a scalar because
the skyrmion is tilted, and its value also depends on the skyrmion size.
However, the asymmetry and the dependence on the skyrmion size is weak.
For the parameter we used in Fig. \ref{move}, it is good enough to
adopt the $\mathcal{D}$ in the corresponding planar film, as shown
in Figs. \ref{move}(c) and (e), which show that the parallel
velocity components are nearly the same in a planar film and in a tube.
However, the perpendicular component $v_\perp$ is significantly different due to the
curving nature of the tube. This can be understood following the
schematic diagrams in Fig. \ref{fig4}(b): Supposing that each layer is independent
to each other, the $v_\perp$ is the same as that in the planar film. However, for
the layer at different $\rho$, the angular speed $\omega(\rho)=v_\perp/\rho$ is
larger (smaller) for smaller (larger) $\rho$. Since the layers are
strongly coupled by exchange interactions and the skyrmion in each layer is closely bounded together,
the smaller angular speed in outer layers is dragged to become faster so that all layers share the same
angular speed in the end. As a result, $v_\perp$ at the outer surface is
faster than that in a planar film, and it increases with $d$ in a nanotube.
The annular speed $\omega=v_\perp/R$ also increases with $d$ for fixed outer
radius $R$. By fitting with the numerical data, we find that $v_\perp$
as well as $\omega$ can be estimated by considering the skyrmion motion
at the layer with radius $R-d/2$, i.e., the half-thickness of the tube wall. At $\rho=R-d/2$,
the Thiele's equation gives
$v_\perp|_{\rho=R-d/2}=\frac{(\alpha-\beta)G\mathcal{D}}{G^2+(\alpha\mathcal{D})^2}v_s=\omega(R-d/2)$. Thus, we obtain
\begin{equation}\label{tubehalf}
v_\perp|_{\rho=R}=\frac{(\alpha-\beta)G\mathcal{D}R}{\left[G^2+(\alpha\mathcal{D})^2\right](R-d/2)}v_s.
\end{equation}
In Figs. \ref{move}(d) and (f), the above expression for $v_\perp$
is plotted in dashed curves, which show excellent agreement with the numerical results.

Magnetic nanotube is the key to test our theoretical predictions. Experimentally, there are several methods to produce the nanotube geometry. Sui \textit{et al.} reported that the nanotubes can be generated by hydrogen reduction in nanochannels of porous alumina templates.\cite{Sui2004} Nielsch \textit{et al.} proposed that the nanotubes can be made by electrodeposition.\cite{Nielsch2005} Daub \textit{et al.} reported that the magnetic nanotubes can be synthesized by atomic layer deposition into porous membranes.\cite{Daub2007}

To conclude, we investigate the static and dynamic properties of skyrmions on magnetic nanotubes. Through micromagnetic simulations, we show that the electric current can drive a skyrmion propagation with a helical trajectory on the tube because of the skyrmion Hall effect. The skyrmion velocity is proportional to the injected current without conventional upper limit. The skyrmion's annular velocity increases with the thickness of the nanotube, which is different from the fact in the planar geometry. Our proposal of transporting the skyrmion in nanotube geometry will stimulate future design of skyrmionic devices.

\begin{acknowledgments}
This work is supported by the National
Natural Science Foundation of China (NSFC) (Grants No. 11604041 and 11704060), the
National Key Research Development Program under Contract No. 2016YFA0300801,
and the National Thousand-Young-Talent Program of China. XSW acknowledges the
support from NSFC (Grant No. 11804045)
and China Postdoctoral Science Foundation (Grant No. 2017M612932 and 2018T110957).
\end{acknowledgments}

\end{document}